\documentclass[aps,twocolumn]{revtex4}

\usepackage{graphics}
\usepackage{epsfig}

\begin{document}
\title{Multi-messenger observations of neutron rich matter}

\author{C. J. Horowitz}

\email{horowit@indiana.edu}
\affiliation{Center for the Exploration of Energy and Matter and Department of Physics, Indiana University, Bloomington, IN 47405, USA}
\date{\today}
\begin{abstract}  
At very high densities, electrons react with protons to form neutron rich matter.  This material is central to many fundamental questions in nuclear physics and astrophysics.  Moreover, neutron rich matter is being studied with an extraordinary variety of new tools such as  Facility for Rare Isotope Beams  (FRIB) and the Laser Interferometer Gravitational Wave Observatory (LIGO).  We describe the Lead Radius Experiment (PREX) that uses parity violating electron scattering to measure the neutron radius in $^{208}$Pb.  This has important implications for neutron stars and their crusts. We discuss X-ray observations of neutron star radii.  These also have important implications for neutron rich matter.  Gravitational waves (GW) open a new window on neutron rich matter.  They come from sources such as neutron star mergers, rotating neutron star mountains, and collective r-mode oscillations.  Using large scale molecular dynamics simulations, we find neutron star crust to be very strong.  It can support mountains on rotating neutron stars large enough to generate detectable gravitational waves.  Finally, neutrinos from core collapse supernovae (SN) provide another, qualitatively different probe of neutron rich matter.  Neutrinos escape from the surface of last scattering known as the neutrino-sphere.  This is a low density warm gas of neutron rich matter.  Neutrino-sphere conditions can be simulated in the laboratory with heavy ion collisions.  Observations of neutrinos can probe nucleosyntheses in SN. Simulations of SN depend on the equation of state (EOS) of neutron rich matter.  We discuss a new EOS based on virial and relativistic mean field calculations.  We believe that combing astronomical observations using photons, GW, and neutrinos, with laboratory experiments on nuclei, heavy ion collisions, and radioactive beams will fundamentally advance our knowledge of compact objects in the heavens, the dense phases of QCD, the origin of the elements, and of neutron rich matter.

\end{abstract}
\maketitle

\section{Introduction}

Multi-messenger astronomy observes matter under extreme conditions.  In this paper we describe how electromagnetic, gravitational wave, and neutrino astronomy, along with laboratory experiments, provide complimentary information on neutron rich matter.  Compress almost anything to very high densities and electrons react with protons to form neutron rich matter.  This material is at the heart of many fundamental questions in Nuclear Physics and Astrophysics.
\begin{itemize}
\item What are the high density phases of QCD?
\item Where did the chemical elements come from?
\item What is the structure of many compact and energetic
objects in the heavens, and what determines their
electromagnetic, neutrino, and gravitational-wave
radiations?
\end{itemize}
Furthermore, neutron rich matter is being studied with an extraordinary
variety of new tools such as the Facility for Rare Isotope
Beams (FRIB), a heavy ion accelerator to be built at
Michigan State University \cite{frib}, and the Laser Interferometer
Gravitational Wave Observatory (LIGO) \cite{ligo}.  Indeed there are many, qualitatively different, probes of neutron rich matter including precision laboratory measurements on stable nuclei and experiments with neutron rich radioactive beams.  While astrophysical observations probe neutron rich matter with electromagnetic radiation, neutrinos, and gravitational waves.   In this paper we give brief examples of how neutron rich matter is being studied with these extraordinarily different probes.

\begin{figure}[h]
\center\includegraphics[width=3.25in]{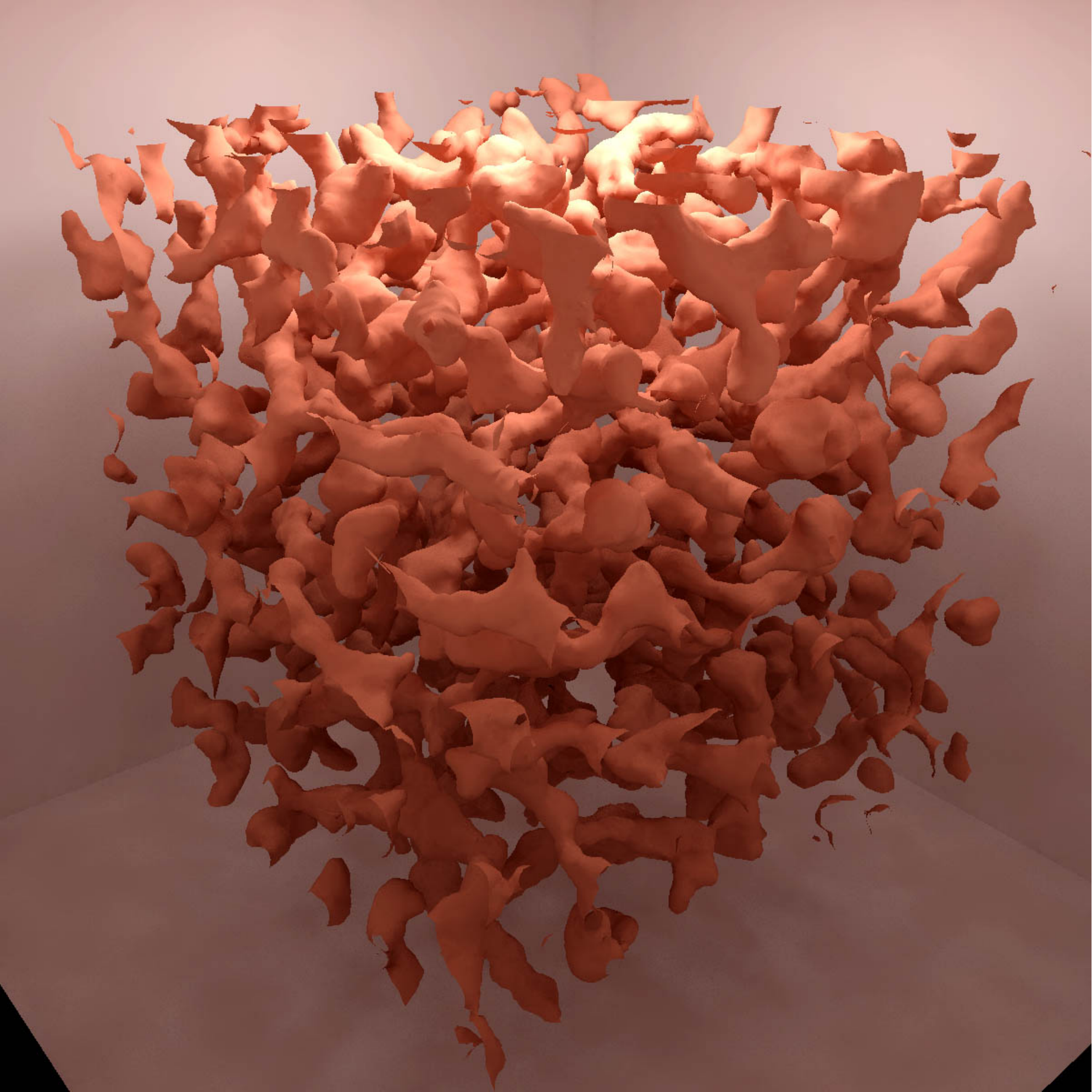}
\caption{\label{Fig1}Surfaces of proton density for a pasta configuration of neutron rich matter at a baryon density of 0.05 fm$^{-3}$.  This is from a semiclassical molecular dynamics simulation with 100,000 nucleons \cite{pasta2}.} 
\end{figure}

We are interested in neutron rich matter over a tremendous range of densities and temperatures were it can be a gas, a liquid, a solid, a plasma, a liquid crystal, a superconductor, a superfluid, a color superconductor, etc.  Neutron rich matter is a remarkably versatile material.  The liquid crystal phases are known as nuclear pasta and arise because of coulomb frustration \cite{pasta}.  Pasta is expected at the base of the crust in a neutron star and can involve complex shapes such as long rods (``spaghetti'') or flat plates (``lasagna'').  Neutrinos in core collapse supernovae may scatter coherently from these shapes (neutrino pasta scattering) because the shapes have sizes comparable to the neutrino wavelength \cite{pastascattering}.  An example of nuclear pasta is shown in Fig. \ref{Fig1}.  

Neutron rich matter is likely a superfluid with a critical temperature of order 10$^{10}$ K.  This superfluid is thought to be important for glitches that are observed in the rotational periods of many pulsars \cite{glitches}.  Here some of the star's angular momentum is carried by rotational vortices in the superfluid.  Sometimes these vortices can move all at once and produce a sudden change in the star's moment of inertia.    

At very high densities, neutron rich matter should form exotic quark and gluon phases.  Asymptotic freedom in QCD implies that at high densities, quarks will be nearly free.  Under these conditions, attractive gluon exchange interactions should pair the quarks into a color superconductor \cite{colorsuperconductor}.  Multi-messenger astronomy is, at present, the only way to observe cold dense matter and search for a color superconductor.  It can not be created in the laboratory.  Although heavy ion collisions can produce high densities, there is no way to get the entropy out and produce low temperatures.   Thus hot dense matter can be studied in the laboratory, but not cold dense matter.  Indeed a very interesting strongly interacting quark gluon plasma has been observed at the Relativistic Heavy Ion Collider (RHIC) \cite{RHIC}.  This very hot material forms a nearly perfect fluid with a low shear viscosity.

In this paper we focus on some of the simpler gas, solid, and liquid phases of neutron rich matter.    In Section \ref{sec.nuclei} we describe a precision laboratory experiment called PREX to measure the neutron radius of $^{208}$Pb.  Nuclei are liquid drops, so PREX and many other laboratory experiments probe the {\it liquid} phase of neutron rich matter.   In astrophysics, electromagnetic, gravitational wave, and neutrino probes can observe different phases of neutron rich matter because the probes have very different mean free paths.  In Section \ref{sec.EM} we describe electromagnetic observations of neutron star radii and crust cooling.  In Section \ref{sec.GW} we discuss gravitational waves from neutron star mergers that are produced by the energetic motions of dense {\it liquid} phase neutron rich matter.   In addition, continuous gravitational waves can be produced by  ``mountains'' of {\it solid} neutron rich matter on rapidly rotating stars.  Finally, as discussed in Section \ref{sec.nu},  neutrinos from core collapse supernovae are radiated from the neutrino-sphere.  This region of last scattering is a low density warm {\it gas}  of neutron rich matter.  Neutrino-sphere conditions can be simulated in the laboratory with heavy ion collisions.  We conclude in Section \ref{sec.conclusions}.

\section{Laboratory probes of neutron rich matter}
\label{sec.nuclei}
Neutron rich matter can be studied in the laboratory.  Hot and or dense matter can be formed in heavy ion collisions, while more neutron rich conditions can be accessed with radioactive beams.  In addition precise experiments are possible on stable neutron rich nuclei.  We give one example, the Lead Radius Experiment (PREX) \cite{prex} accurately measures the neutron radius in $^{208}$Pb with parity violating electron scattering \cite{bigprex}.  This has many implications for nuclear structure, astrophysics, atomic parity violation, and low energy tests of the standard model. 

\subsection{Introduction to neutron densities and neutron radii}  

Nuclear charge densities have been accurately measured with electron scattering and have become our picture of the atomic nucleus, see for example ref. \cite{chargeden}.  These measurements have had an enormous impact.  
In contrast, our knowledge of neutron densities comes primarily from hadron scattering experiments involving for example pions \cite{pions}, protons \cite{protons1,protons2,protons3}, or antiprotons \cite{antiprotons1,antiprotons2}.  However, the interpretation of hadron scattering experiments is model dependent because of uncertainties in the strong interactions.

Parity violating electron scattering provides a model independent probe of neutron densities that is free from most strong interaction uncertainties.  This is because the weak charge of a neutron is much larger than that of a proton \cite{dds}.  Therefore the $Z^0$ boson, that carries the weak force, couples primarily to neutrons.  In Born approximation, the parity violating asymmetry $A_{pv}$, the fractional difference in cross sections for positive and negative helicity electrons, is proportional to the weak form factor.  This is very close to the Fourier transform of the neutron density.  Therefore the neutron density can be extracted from an electro-weak measurement \cite{dds}.  
Many details of a practical parity violating experiment to measure neutron densities have been discussed in a long paper \cite{bigprex}.    

\begin{figure}[h]
\center\includegraphics[width=3.25in]{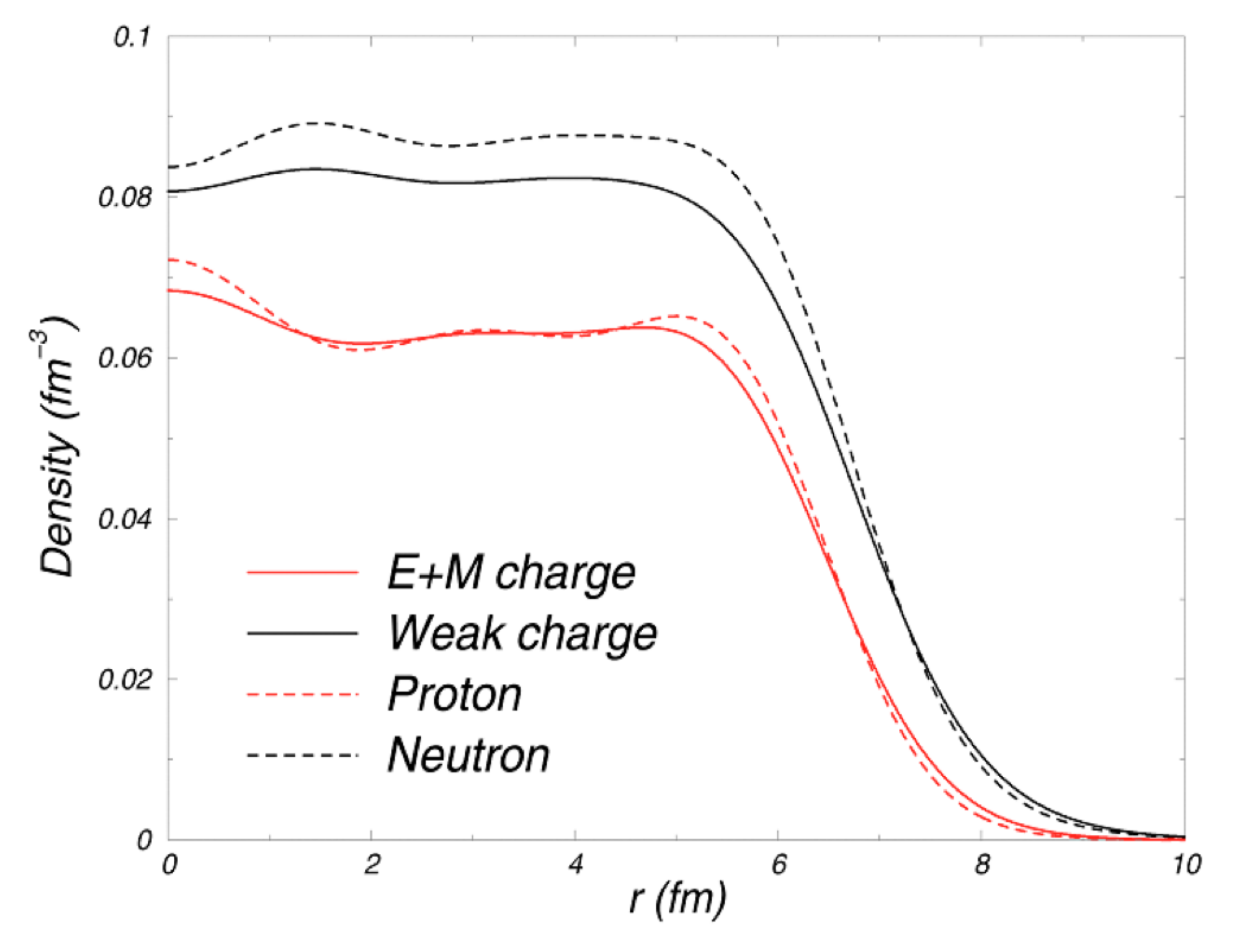}
\caption{\label{Fig2}Densities of $^{208}$Pb in a relativistic mean field model \cite{cjh_bds}.  The lower dashed (red) curve shows the point proton density while the upper dashed (black) curve is the neutron density.  The (electromagnetic) charge density is the lower solid (red) curve while the upper solid (black) curve shows the weak charge density that is measured in the PREX experiment.}
\end{figure}

The doubly magic nucleus $^{208}$Pb has 44 more neutrons than protons, and some of these extra neutrons are expected to be found in the surface where they form a neutron rich skin, see Fig. \ref{Fig2}.  The thickness of this skin is sensitive to nuclear dynamics and provides fundamental nuclear structure information.  There may be a useful analogy with cold atoms in laboratory traps were similar ``spin skins'' have been observed for partially polarized systems  \cite{coldatomsMIT, coldatomsRICE}.  Note that there is an attractive interaction between two atoms of unlike spins while for a nucleus, the interaction between two nucleons of unlike isospins is also more attractive than the interaction between two nucleons of like isospins.       

The neutron radius of $^{208}$Pb, $R_n$, has important implications for astrophysics.  There is a strong correlation between $R_n$ and the pressure of neutron matter $P$ at densities near 0.1 fm$^{-3}$ (about 2/3 of nuclear density) \cite{alexbrown}.  A larger $P$ will push neutrons out against surface tension and increase $R_n$.  Therefore measuring $R_n$ constrains the equation of state (EOS) --- pressure as a function of density --- of neutron matter.  

Recently Hebeler et al. \cite{hebeler} used chiral perturbation theory to calculate the EOS of neutron matter including important contributions from very interesting three neutron forces.  
From their EOS, they predict $R_n-R_p= 0.17 \pm 0.03$ fm.  Here $R_p$ is the known proton radius of $^{208}$Pb.   Monte Carlo calculations by Carlson et al. also find sensitivity to three neutron forces \cite{MC3n}.   Therefore, measuring $R_n$ provides an important check of fundamental neutron matter calculations, and constrains three neutron forces.

The correlation between $R_n$ and the radius of a neutron star $r_{NS}$ is also very interesting \cite{rNSvsRn}.  In general, a larger $R_n$ implies a stiffer EOS, with a larger pressure, that will also suggest $r_{NS}$ is larger.  Note that this correlation is between objects that differ in size by 18 orders of magnitude from $R_n\approx 5.5$ fm to $r_{NS}\approx 10$ km.  We discuss observations of $r_{NS}$ in Section \ref{sec.EM}. 
 

The EOS of neutron matter is closely related to the symmetry energy $S$.  
This describes how the energy of nuclear matter rises as one goes away from equal numbers of neutrons and protons.  There is a strong correlation between $R_n$ and the density dependence of the symmetry energy $dS/dn$, with $n$ the baryon density.  The symmetry energy can be probed in heavy ion collisions \cite{isospindif}.  For example, $dS/dn$ has been extracted from isospin diffusion data \cite{isospindif2} using a transport model.

The symmetry energy $S$ helps determine the composition of a neutron star.     A large $S$, at high density, implies a large proton fraction $Y_p$ that will allow the direct URCA process of rapid neutrino cooling.  If $R_n-R_p$ is large, it is likely that massive neutron stars will cool quickly by direct URCA  \cite{URCA}.  In addition, the transition density from solid neutron star crust to the liquid interior is strongly correlated with $R_n-R_p$ \cite{cjhjp_prl}.  

Finally, atomic parity violation (APV) is sensitive to $R_n$ \cite{pollockAPV},\cite{brownAPV},\cite{bigprex}.  Parity violation involves the overlap of atomic  electrons with the weak charge of the nucleus, and this is primarily carried by the neutrons.  Furthermore, because of relativistic effects the electronic wave function can vary rapidly over the nucleus.  Therefore, the APV signal depends on where the neutrons are and on $R_n$.   A future low energy test of the standard model may involve the combination of a precise APV experiment along with PV electron scattering to constrain $R_n$.  Alternatively, measuring APV for a range of isotopes can provide information on neutron densities \cite{berkeleyAPV}.

\subsection{Neutron skins, electric polarizability, and pigmy resonances}

The neutron radius in $^{208}$Pb may also be correlated with a number of other nuclear properties.  Recently Reinhard and Nazarewicz have developed a very nice way to study these correlations \cite{Reinhard}.  The perform least squares fits of energy density functionals to a number of nuclear observables and look at correlations between $R_n$ and several other observables.  They find that $R_n$ is strongly correlated with the dipole polarizability $\alpha_D$ of $^{208}$Pb.  Here $\alpha_D$ involves an inverse energy weighted sum over excited states $\Phi_n$ of energy $E_n$,
\begin{equation}
\alpha_D=2\sum_{n>0} \frac{|\langle \Phi_n|\hat D|\Phi_0\rangle |^2}{E_n}\, ,
\label{alpha}
\end{equation}
$\Phi_0$ is the ground state, and $\hat D$ is the electric dipole operator.

Recently Tamii et al. have accurately determined $\alpha_D$ from very small angle proton scattering \cite{Tamii}.  From this measured value, and Reinhard and Nazarewicz's correlation, they make the prediction $R_n-R_p=0.156^{+0.025}_{-0.021}$ fm.  This value is very promising because of the small error bars.  However, we note that it is still model dependent because of the assumed correlation between $R_n$ and $\alpha_D$.

The pigmy resonance is a low energy electric dipole excitation that has been observed in neutron rich nuclei \cite{pigmy}, see for example \cite{pigmy2}.  The pigmy is thought to involve an oscillation of the neutron skin against the rest of the nucleus.  The total dipole strength in the pigmy resonance is somewhat related to $\alpha_D$ because the $1/E_n$ wight factor in Eq. \ref{alpha} emphasizes low energy excitations.  The fraction of the energy weighted sum rule exhusted by the pigmy is sensitive to $R_n$ \cite{pigmyRn,pigmyRn1,pigmyRn2}.  Recent pioneering experiments on unstable neutron rich isotopes of Sn, Sb, and Ni seem to support this assertion \cite{Klimkiewicz,Adrich,Wieland}.  

\subsection{The Lead Radius Experiment (PREX)}

We now discuss a direct measurement of $R_n$.  Parity violation provides a model independent probe of neutrons, because the $Z^0$ boson couples to the weak charge, and the weak charge of a proton $Q_W^p$,
\begin{equation}
Q_W^p=1-4\sin^2\Theta_W \approx 0.05,
\label{qwp}
\end{equation}
is much smaller than the weak charge of a neutron $Q_W^n$,
\begin{equation}
Q_W^n=-1.
\end{equation}
In Eq. \ref{qwp}, $\Theta_W$ is the weak mixing angle.  Therefore, the weak neutral current interaction, at low momentum transfers, couples almost completely to neutrons.

The Lead Radius Experiment (PREX) at Jefferson Laboratory \cite{prex} measures the parity violating asymmetry $A_{pv}$ for elastic electron scattering from $^{208}$Pb.  The asymmetry $A_{pv}$ is the fractional cross section difference for scattering positive (+), or negative (-), helicity electrons,
\begin{equation}
A_{pv}=\frac{\frac{d\sigma}{d\Omega}|_+-\frac{d\sigma}{d\Omega}|_-}{\frac{d\sigma}{d\Omega}|_++\frac{d\sigma}{d\Omega}|_-}\, .
\end{equation}
In Born approximation, $A_{pv}$ arrises from the interference of a weak amplitude of order the Fermi constant $G_F$, and an electromagnetic amplitude of order the fine structure constant $\alpha$ over the square of the momentum transfer $q^2$ \cite{dds},
\begin{equation}
A_{pv}\approx \frac{G_F q^2\, F_W(q^2)}{2 \pi\alpha\sqrt{2} \, F_{ch}(q^2)}\, .
\label{apvborn}
\end{equation}
Here the weak form factor $F_W(q^2)$ is the Fourier  transform of the weak charge density $\rho_W(r)$, that is essentially the neutron density,
\begin{equation}
F_W(q^2)=\int d^3r \frac{\sin(qr)}{qr} \rho_W(r)\, .
\end{equation}
Likewise, the electromagnetic form factor $F_{ch}(q^2)$ is the Fourier transform of the (electromagnetic) charge density $\rho_{ch}(r)$.  This is known from elastic electron scattering \cite{chargeden}. Therefore, in principle, measuring $A_{pv}$ as a function of $q$ allows one to map out the neutron density $\rho_n(r)$.  Note that, for a heavy nucleus, there are important corrections to Eq. \ref{apvborn} from Coulomb distortions.  However, these have been calculated exactly by solving the Dirac equation for an electron moving in both a Coulomb potential of order 25 MeV and a weak axial vector potential of order electron volts \cite{couldist}.  Therefore, even with Coulomb distortions, one can accurately determine neutron densities.   Note that this purely electroweak reaction is free from most strong interaction uncertainties.

The PREX experiment measures $A_{pv}$ for 1.05 GeV electrons elastically scattered from $^{208}$Pb at laboratory angles near five degrees.  The experiment uses a 0.5 mm thick Pb foil target that is enriched in $^{208}$Pb and sandwiched between two thin diamond foils. These diamond foils help remove the beam heating and keep the target from melting.  Scattered electrons are deflected by septum magnets into the two high resolution spectrometers that separate elastically from inelastically scattered electrons.  The goal of this challenging experiment is to measure $A_{pv}$, that is of order 0.5 ppm, to 3\%.   This allows one to determine the neutron root mean square radius $R_n$ to 1\% ($\pm 0.05$ fm).

The experiment ran in Hall A at Jefferson Laboratory during April to June 2010.  Krishna Kumar, one of the PREX spokespersons, provided the following preliminary results \cite{kk}.  The experiment was successfully commissioned in March/ April.  High quality data were accumulated in transverse spin mode so as to make the systematic error from a potential 2 photon exchange amplitude negligible.  However, radiation issues downstream from the main apparatus resulted in a significant reduction in experiment ``up time''.   Sufficient statistics were accumulated to provide a significant first constraint on the neutron radius.  The statistical error bar is of order 3\% ($\pm 0.15$ fm) for $R_n$.  Systematic errors are much smaller.  Although the statistical errors are presently larger than in the original proposal, this is nevertheless a significant achievement.  The PREX experiment has demonstrated, for the first time, the use of parity violation to measure neutron densities.  Furthermore, the achieved accuracy is sufficient to clearly observe the effects of coulomb distortions.   Coulomb distortions are expected to reduce the parity violating asymmetry $A_{pv}$, from that predicted in Eq. \ref{apvborn}, by about 25\%.  Although there may be little theoretical uncertainty in the predictions of coulomb distortions, nevertheless this represents the first time that their effects on $A_{pv}$ have been directly observed.

Finally, the collaboration has achieved the desired small systematic errors.  Therefore the measurement can be improved simply by accumulating more statistics.  The collaboration purposes to design appropriate engineering modifications to the beamline to mitigate the radiation problem and has been granted additional beam time to improve the statistics and achieve the original goal of a 1\% ($\pm 0.05$ fm) constraint on the neutron radius of $^{208}$Pb.

In addition to PREX, many other parity violating measurements of neutron densities are possible, see for example \cite{PREXII}.  Neutron radii can be measured in many stable nuclei, as long as the first excited state is not too low in energy (so that elastically scattered electrons can be separated from inelastic excitations).  In general, $R_n$ is easier to measure in lighter neutron rich nuclei.  This is because $R_n$ is smaller and so can be measured at higher momentum transfers where $A_{pv}$ is larger.   

Measuring $R_n$ in $^{48}$Ca is particularly attractive.  First, $^{48}$Ca has a higher experimental figure of merit than $^{208}$Pb.  Therefore a $^{48}$Ca measurement may take less beam time than for $^{208}$Pb.  Not only does $^{48}$Ca have a large neutron excess, it is also relatively light.  With only 48 nucleons, microscopic coupled cluster calculations \cite{coupledcluster}, or no core shell model calculations \cite{NCSM}, may be feasible for $^{48}$Ca that are presently not feasible for $^{208}$Pb.   Note that these microscopic calculations may have important contributions from three nucleon forces.  This will allow one to make microscopic predictions for the neutron density and relate a measured $R_n$ to three nucleon forces and in particular to very interesting three neutron forces.  

We end this section with a discussion of perhaps the ultimate neutron density measurement.  Measuring $A_{pv}$ for a range of momentum transfers $q$, see Eq. \ref{apvborn}, allows one to directly determine the neutron density $\rho_n(r)$ as a function of radius $r$.   For $^{208}$Pb this may be extraordinarily difficult.  However, for $^{48}$Ca this may actually be feasible, although somewhat difficult and time consuming.  For example, one might be able to determine about 6 Fourier Bessel coefficients in an expansion of $\rho_n(r)$.  Note that this may require long runs and careful control of backgrounds and could be helped by using a large acceptance spectrometer.  Nevertheless, one should not underestimate the utility of having model independent determinations of both the neutron and proton densities as a function of $r$.  This will literally provide our picture of the neutrons and protons inside an atomic nucleus.

\section{Electromagnetic observations of neutron stars}
\label{sec.EM}

Neutron stars are collapsed stellar objects that are formed in supernova explosions.  They are more massive than the sun but have radii of order 10 km.  This makes them an extraordinary 18 orders of magnitude larger than a $^{208}$Pb nucleus and 55 orders of magnitude more massive.
Nevertheless, both in the laboratory and in astrophysics, {\it it is the same neutrons, the same strong interactions, the same neutron rich matter, and the same equation of state}.  A measurement in one domain, be it astrophysics or in the laboratory, can have important implications in the other domain.  This is the real strength of a multi-messenger approach to neutron rich matter. 

\subsection{Neutron star radii}
The structure of a neutron star can be calculated with the Tolman-Oppenheimer-Volkoff Equations of General Relativity \cite{TOV} and is completely determined by the equation of state of neutron rich matter.  The radius of a neutron star depends on the pressure of neutron matter at normal nuclear density and above, because the central density of a neutron star can be a few or more times that of normal nuclear density.   A higher pressure will lead to a larger radius.  It is important to have both low density information on the equation of state from PREX, and high density information from measurements of neutron star radii.  This can constrain any possible density dependence of the equation of state from an interesting phase transition to a possible high density exotic phase such as quark matter, strange matter, or a color superconductor.  For example, if the $^{208}$Pb radius is relatively large, this shows the EOS is stiff at low density (has a high pressure).  If at the same time, neutron stars have relatively small radii, than the high density EOS is soft with a low pressure.  This softening of the EOS with density could strongly suggest a phase transition to a soft high density exotic phase.

The radius of a neutron star $r_{NS}$ can be deduced from X-ray measurements of luminosity $L$ and surface temperature $T$,
\begin{equation}
L=4\pi r_{NS}^2 \sigma_{SB}T^4,
\label{NSR}
\end{equation}       
with $\sigma_{SB}$ the Stefan Boltzmann constant.  This allows one to deduce the surface area of a neutron star.  However, there are important complications.  First, one needs an accurate distance to the neutron star.   Second, Eq. \ref{NSR} assumes a black-body and there are important non-blackbody corrections that must be determined from models of neutron star atmospheres.  Finally, gravity is so strong that space near a neutron star is strongly curved.  If one looks at the front of a neutron star, one will also see about 30\% of the back, because of the curvature of space.  Therefore the surface area that one observes in Eq. \ref{NSR} depends on the curvature of space and the mass of the star.  Thus what started out as a measurement of just the radius is, in fact, a measurement of a combination of mass and radius. 

Recently Steiner, Lattimer, and Brown have deduced masses and radii \cite{SLB} from combined observations of six neutron stars in two classes: 1) X-ray bursts, and 2) neutron stars in globular clusters.  They conclude that observations favor a stiff high density equation of state that can support neutron stars with a maximum mass near 2 $M_\odot$ and that the equation of state is soft at low densities so that a 1.4 $M_\odot$ neutron star has a radius near 12 km.  They go on to predict that the neutron minus proton root mean square radius in $^{208}$Pb will be $R_n-R_p=0.15\pm 0.02$ fm.  Note that this is a prediction for a nucleus based on an equation of state deduced from X-ray observations of neutron stars.

The Steiner et al. paper \cite{SLB} is potentially controversial because their results depend on, among other things, the model assumed for X-ray bursts.  Ozel et al. use a different model for X-ray bursts,  and get very small neutron star radii near 10 km or below \cite{ozel}.   The high density EOS that Ozel et al. deduce from these observations is significantly softer than Steiner et al's, suggesting a phase transition to an exotic phase \cite{ozelEOS}.  Thus, the difference between Ozel et al's result of 10 km, that likely implies a phase transition, and Steiner et al's result of 12 km, that does not suggest a phase transition, is very significant.  Clearly these observations of neutron star radii have potentially very important implications for the properties of neutron rich matter and the high density phases of QCD.

\subsection{Solid neutron rich matter and neutron star crust cooling}

We go on to discuss the neutron star crust and crust cooling.  The neutron radius of $^{208}$Pb has implications for neutron star structure, in addition to the star's radius.   Neutron stars have solid crusts over liquid cores, while a heavy nucleus is expected to have a neutron rich skin.  Neutron star crust consists of a relativistic fermi gas of electrons, a crystal lattice of neutron rich ions, and, in general, a neutron gas.  For a review see \cite{crustreview}.  Both the solid crust of the star, and the skin of the nucleus, are made of neutron rich matter at similar, slightly subnuclear, densities.   The common unknown is the equation of state of neutron matter.   A thick neutron skin in $^{208}$Pb, means a high pressure where the energy rises rapidly with density.  This quickly favors the transition to a uniform liquid phase.  Therefore, we find a strong correlation between the neutron skin thickness, measured by PREX, and the transition density in neutron stars from solid crust to liquid interior \cite{cjhjp_prl}.     

Perhaps the presence of a solid crust deserves comment.  We expect solids to form in compact stars, both white dwarfs and neutron stars.  This may sound surprising since stars are often composed of not liquids or gasses, but plasmas.  Nevertheless, the plasmas can be so dense that the ions actually freeze.  Recently, there are significant new observations of how white dwarfs freeze \cite{winget} and of how the solid crust of accreting neutron stars cools \cite{crustcooling,crustcooling1,crustcooling2}.   In addition, crystalization has been observed in complex laboratory plasmas, see for example \cite{complexplasmas,complexplasmas2}.

\subsubsection{White dwarf crystallization}
We start with white dwarf crystallization.  Winget et al. deduce the luminosity function for white dwarfs in the globular star cluster NGC 6397 \cite{winget}, see also \cite{garcia-berro}.  This is simply the number of white dwarfs in different luminosity bins.  As white dwarfs cool, their luminosities decrease.  The probability to find a white dwarf with a given luminosity depends on the cooling rate.  One is more likely to find stars at luminosities where they cool slowly, and hence spend a long time.  Therefore, the luminosity function is expected to be proportional to one over the cooling rate.   Eventually as a star cools, it freezes.  When this happens, the latent heat will slow the cooling rate until finally all of the latent heat is radiated away.  This is exactly like ice cubes slowing the rate of warming of your drink.  Therefore there should be a peak in the luminosity function when the stars freeze.  Indeed such a peak is observed by Winget et al. \cite{winget}.  Furthermore, the luminosity where this peak occurs, depends on the interior temperature.  This allows one to deduce the melting temperature of white dwarf cores.   Comparing this melting temperature to the phase diagram of carbon/ oxygen mixtures determined from large scale molecular dynamics simulations allows one to set limits on the ratio of carbon to oxygen in these white dwarf cores, and the rate of the very interesting but incompletely known, nuclear reaction $^{12}$C($\alpha,\gamma$)$^{16}$O \cite{wdprl}.  

\subsubsection{Crystallization in accreting neutron stars}
We next consider freezing in accreting neutron stars.  Material falling on a neutron star can undergo rapid proton capture (or rp process) nucleosynthesis to produce a range of nuclei with mass numbers $A$ that could be as high as $A\approx 100$ \cite{rpash, rpash2}.  As this rp process ash is buried by further accretion, the rising electron fermi energy induces electron capture to produce a range of neutron rich nuclei from O to approximately Se \cite{rpash3}.  This material freezes, when the density reaches near $10^{10}$ g/cm$^3$.  We have performed large scale MD simulations of how this complex rp process ash freezes \cite{phasesep}.  We find that chemical separation takes place and the liquid ocean is greatly enriched in low atomic number $Z$ elements, while the newly formed solid crust is enriched in high $Z$ elements.  Furthermore, we find that a regular crystal lattice forms even though large numbers of impurities are present.  This regular crystal should have a high thermal conductivity.  We do not find an amorphous solid that would have a low thermal conductivity \cite{soliddiffusion}.

Recent X-ray observations of neutron stars, find that the crust cools quickly, when heating from extended periods of accretion stops \cite{crustcooling,crustcooling1,crustcooling2}.  This is consistent with our MD simulations, and strongly favors a crystalline crust over an amorphous solid that would cool more slowly \cite{crustcooling3},  \cite{crustcooling4}.  In Sec. \ref{sec.GW}, we will be interested in the strength of this crystalline crust.

\section{Gravitational Waves}
\label{sec.GW}

We turn now to gravitational wave observations of neutron rich matter.   Albert Einstein, almost 100 years ago, predicted the oscillation of space and time known as gravitational waves (GW).  Within a few years, with the operation of Advanced LIGO \cite{advancedLIGO}, Advanced VIRGO \cite{advancedVIRGO} and other sensitive interferometers, we anticipate the historic detection of GW.  This will be a remarkable achievement and open a new window on the universe and on neutron rich matter.  

The first GW that are detected will likely come from the merger of two neutron stars.  The rate of such mergers can be estimated from known binary systems \cite{LIGOrate}.   During a merger the GW signal has a so called chirp form where the frequency rises as the two neutron stars spiral closer together.  Deviations of this wave form from that expected for two point masses may allow one to deduce the equation of state of neutron rich matter and measure the radius of a neutron star $r_{NS}$ \cite{GWEOS}.   Alternatively one may be able to observe the frequency of oscillations of the hyper-massive neutron star just before it collapses to a black hole.  This frequency depends on the radius of the maximum mass neutron star \cite{Rmax}.  However, either approach may require high signal to noise data from relatively nearby mergers.  Note that this information on $r_{NS}$ from GW is independent of the X-ray observations discussed in Sec. \ref{sec.EM} and of any systematic errors associated with the modeling of X-ray bursts.     

Continuous GW signals can also be detected, see for example \cite{Collaboration:2009rfa}, in addition to burst signals such as those from neutron star mergers.  In this colloquium, we will focus on continuous gravitational waves.  Indeed Bildstein and others \cite{bildstein} have speculated that some neutron stars in binary systems may radiate angular momentum in continuous GW at the same rate that angular momentum is gained from accretion.  This would explain why the fastest observed neutron stars are only spinning at about half of the breakup rate.  There are several very active ongoing and near future searches for continuous gravitational waves at LIGO, VIRGO and other detectors, see for example \cite{abbot}.  Often one searches at twice the frequency of known radio signals from pulsars because of the quadrupole nature of GW.  No signal has yet been detected.  However, sensitive upper limits have been set.  These limits constrain the shape of neutron stars.  In some cases the star's elipticity $\epsilon$, which is that fractional difference in moments of inertia $\epsilon=(I_1-I_2)/I_3$ is observed to be less than a part per million or even smaller.  Here $I_1$, $I_2$, and $I_3$ are the principle moments of inertia.    

In general, the amplitude of any continuous signal is much weaker than a burst signal.  However, one can gain sensitivity to a continuous signal by coherently (or semi-coherently) integrating over a large observation time, see for example \cite{semicoherent}.  Note that searches for continuous GW can be very computationally intensive because one must search over an extremely large space of parameters that may include the source frequency, how that frequency changes with time, the source location on the sky, etc.  The Einstein at home distributed computing project uses spare cycles on the computers of a large number of volunteers to search for continuous GW \cite{Einstein_home}. 
           
Strong GW sources often involve large accelerations of large amounts of neutron rich matter.  Indeed the requirements for a strong source of continuous GW, at LIGO frequencies, places extraordinary demands on neutron rich matter.   Generating GW sounds easy.  Place a mass on a stick and shake vigorously.  However to have a detectable source, one may need not only a large mass, but also a very strong stick.   The stick is needed to help produce large accelerations.  Since others have discussed large masses, let us focus here on the strong stick.

An asymmetric mass on a rapidly rotating neutron star produces a time dependent mass quadrupole moment that radiates gravitational waves.  However, one needs a way (strong stick) to hold the mass up.  Magnetic fields can support mountains, see for example \cite{magneticmountains}.  However, it may require large internal magnetic fields.  Furthermore, if a star also has a large external dipole field, electromagnetic radiation may rapidly spin the star down and reduce the GW radiation.  

\begin{figure}[h]
\center\includegraphics[width=3.25in]{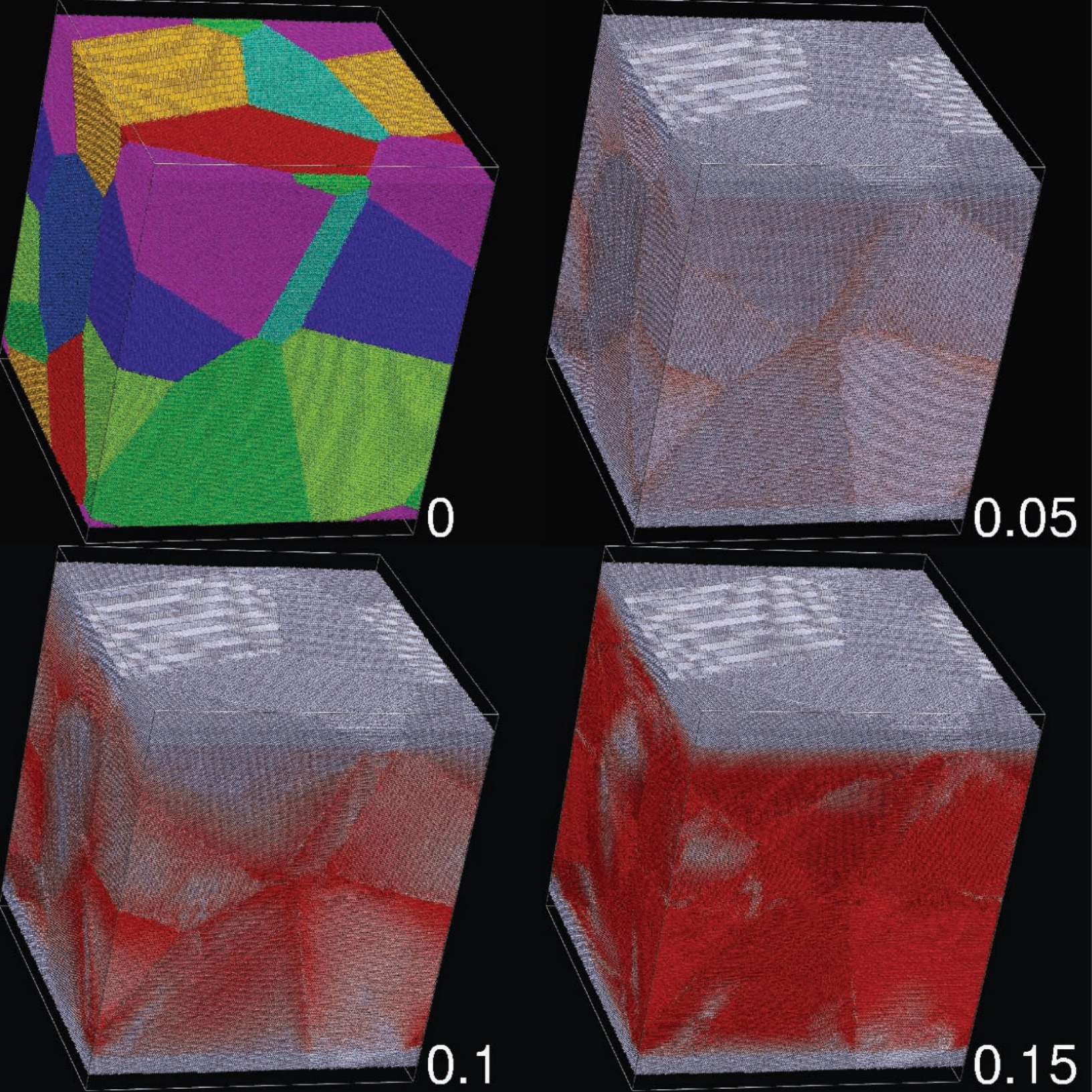}
\caption{\label{Fig3}Shear deformation of a 12 million ion polycrystaline sample of neutron star crust.  The panels are labeled by the strain, which is the fractional deformation.   The colors in the upper left panel, at zero strain, show the original eight microcrystals of different orientations.  The red color in the other three panels indicates distortion of the body centered cubic lattice \cite{crustbreaking}.}
\end{figure}

Alternatively, mountains can be supported by the solid neutron star crust.
Recently we performed large scale MD simulations of the strength of neutron star crust \cite{crustbreaking,chugunov}.   A strong crust can support large deformations or ``mountains'' on neutron stars, see also \cite{lowmassNS}, that will radiate strong GW.   How large can a neutron star mountain be before it collapses under the extreme gravity?  This depends on the strength of the crust.  We performed large scale MD simulations of crust breaking, where a sample was strained by moving top and bottom layers of frozen ions in opposite directions \cite{crustbreaking}.  These simulations involve up to 12 million ions and explore the effects of defects, impurities, and grain boundaries on the breaking stress.  For example, in Fig. \ref{Fig3} we show a polycrystalline sample involving 12 million ions.  In the upper right panel the initial system is shown, with the different colors indicating the eight original microcrystals that make up the sample.  The other panels are labeled with the strain, i.e. fractional deformation, of the system.  The red color indicates distortion of the body centered cubic crystal lattice.  The system starts to break along grain boundaries.  However the large pressure holds the microcrystals together and the system does not fail until large regions are deformed.

We find that neutron star crust is very strong because the high pressure prevents the formation of voids or fractures and because the long range coulomb interactions insure many redundant ``bounds'' between planes of ions.  Neutron star crust is the strongest material known, according to our simulations.  The breaking stress is 10 billion times larger than that for steel.  This is very promising for GW searches because it shows that large mountains are possible, and these could produce detectable signals.  

Continuous GW can also be produced by r-mode oscillations of a rotating neutron star \cite{r-modereview}.  Consider a surface wave on a rapidly rotating neutron star that is moving slowly in a direction opposite to the stars rotation.  This wave, in the laboratory frame, will appear to be moving in the direction of the star's rotation.  The back reaction force on the wave from GW radiation will always act to slow the wave in the laboratory frame.  However, in this case slowing in the lab frame will actually speed up the wave in the rotating star's frame and increase its amplitude.  Thus the wave can be unstable with respect to gravitational wave radiation.   The r-modes are collective oscillations on rotating neutron stars that can also be unstable to GW radiation \cite{r-modereview}.  If an r-mode is unstable the amplitude will grow large and rotational kinetic energy can be radiated away as GW.  This will slow the rotation rate.  Note, that the physics of large amplitude r-mode oscillations can be complicated, see for example \cite{largermode,largermode2}.

The stability of r-modes depends on the amount of dissipation from, for example, the bulk and shear viscosities of neutron rich matter.  If dissipation is large then the amplitude of the r-modes will stay small and rapid rotation of a neutron star is possible.  Alternatively, if dissipation is small then the r-modes may be unstable and GW radiation from the modes may limit the rotation rate of the star.  Unfortunately the stability of r-modes has proved to be a complex subject that may be sensitive to subtle dissipation properties of neutron rich matter, be it in a nucleon phase \cite{shearviscosity1,shearviscosity2} or in more exotic quark and gluon phases \cite{kaonbulk, quarkbulk, quarkviscosity}.

We give one example of a possible source of dissipation for the r-modes.  The shear viscosity of conventional complex fluids, with large non-spherical molecules, can be orders of magnitude larger than that for normal fluids.  This suggests that the shear viscosity of nuclear pasta, with long rod like shapes as seen in Fig. \ref{Fig1}, could be large.   In ref. \cite{pastaviscosity}, we have calculated the shear viscosity of nuclear pasta using large scale molecular dynamics simulations.  The shear viscosity is dominated by momentum carried by electrons and although the electron mean free path is determined by electron-pasta scattering, we find no dramatic differences from a conventional phase with spherical nuclei.  Therefore, we find that the shear viscosity of nuclear pasta is not very different from that for more conventional matter with nearly spherical nuclei.  However there could be other sources of dissipation.  We do not yet have a complete understanding of when the r-modes may be stable and when they are unstable.  

To conclude this section, there is a great deal of interest in gravitational waves (GW) from neutron stars and there are many ongoing searches.  One is interested in both burst sources, for example from neutron star mergers, and  continuous sources from mountains or collective modes.  Gravitational wave radiation depends on the equation of state of neutron rich matter.  In addition, it can also depend on other more detailed properties including the breaking strain of solid phases and the bulk and shear viscosities.              

\section{Supernova neutrinos and neutron rich matter}
\label{sec.nu}

Neutrinos provide yet another, qualitatively different, probe of neutron rich matter.  Core collapse supernovae (SN) are gigantic stellar explosions that convert as much as $0.2 M_\odot$ of mass into $10^{58}$ neutrinos \cite{SNreview}.   We detected about 20 neutrino events from SN1987a \cite{SN1987a, SN1987ab}.  The next galactic SN should be very exciting.  There are many new underground experiments to search for dark matter, double beta decay, solar neutrinos, proton decay, oscillation of accelerator neutrinos, etc.   Some of these experiments can be very sensitive to SN neutrinos.  For example, the coming ton scale dark matter experiments can be sensitive to SN neutrinos via neutrino-nucleus elastic scattering \cite{mckinsey}.  Here, the energy of recoiling nuclei provides information on the spectrum of mu and tau neutrinos from a SN.  This information is not available from conventional water detectors such as Super-Kamiokande \cite{SuperK}.  Super-K is primarily sensitive to $\bar\nu_e$ via $\bar\nu_e+p\rightarrow n + e^+$.  Note that most of the mass of Super-K is oxygen and this is less sensitive to SN neutrinos than hydrogen.  In contrast, a neutrino-nucleus elastic detector involves very large coherent cross sections, is sensitive to all six flavors of SN neutrinos ($\nu_e,\bar\nu_e,\nu_\mu,\bar\nu_\mu,\nu_\tau$, and $\bar\nu_\tau$) and all of the mass, in the fiducial volume, is active.  This gives very large yields of tens to hundreds of events per {\it ton}, for a SN at 10 kiloparsecs \cite{mckinsey}.  This should be compared to yields of hundreds of events per {\it kiloton} for conventional SN detectors.            

In a SN, neutrinos are emitted from the surface of last scattering known as the neutrino-sphere.  At the neutrino-sphere, the neutrino mean free path is comparable to the size of the system.  The temperature is of order 4 MeV, based on the energies of the SN1987a events. The density is $\approx 10^{11}$ g/cm$^3$, 1/1000 to 1/100 of normal nuclear density.   This follows from the mean free path implied by known neutrino cross sections.    Thus, the neutrino-sphere is a low density warm gas of neutron rich matter.  The pressure, composition, and long wave-length neutrino response \cite{nuresponse} of this region can be calculated from a model independent Virial expansion \cite{virial,virialnuc}.   This expansion is based on nucleon-nucleon, nucleon-alpha, and alpha-alpha elastic scattering phase shifts.    

Neutrino-sphere conditions can be simulated in the laboratory with heavy ion (HI) collisions.  For example during a peripheral collision, fragments emitted with velocities intermediate between those of the projectile and target are coming from a warm low density region, see for example \cite{Natowitz}.  
These experiments studied the composition of light nuclei, for near neutrino-sphere conditions, and there appears to be good agreement with Virial predictions \cite{Natowitz}.   Note that the abundance of light nuclei such as $^3$He or $^3$T, near the neutrino-sphere, can be important for the radiated $\nu_e$ spectrum \cite{mass3}, see below.  Temperatures in these HI collisions are measured to be directly comparable to temperatures of the neutrino-sphere.  The density is low, significantly below nuclear density.  However, methods to measure densities during HI collisions may not be as well developed as methods to measure temperatures.  Finally, present HI collisions are not as neutron rich as is the neutrino-sphere.  However, future experiments can involve more neutron rich radioactive beams.

\subsection{r-process nucleosynthesis, supernova neutrinos, and the search for Eldorado}
The next galactic SN should provide a great deal of information on neutrino properties such as oscillations, masses, and mixing angles.  We give one example of an important SN neutrino observable related to nucleosynthesis.   About half of the elements heavier than $^{56}$Fe, including gold and uranium, are thought to be made in the r-process of rapid neutron capture nucleosynthesis \cite{rprocess}.  Here seed nuclei rapidly capture many neutrons to produce very neutron rich nuclei that then beta decay several times to produce heavy elements such as gold.  At this time, the preferred site for the r-process is the neutrino driven wind during a supernova, see for example \cite{rwind}.  Here some baryons are blown off of the proto-neutron star by the intense neutrino flux.  Nucleosynthesis in this wind depends on the entropy, the expansion time scale, and most importantly the proton fraction $Y_p$ (number of protons divided by the total number of protons and neutrons).  The proton fraction is set by the relative rates of neutrino and antineutrino capture.  
\begin{equation}
\nu_e+n\rightarrow p + e
\end{equation}
\begin{equation}
\bar\nu_e+p\rightarrow n + e^+
\end{equation}
The cross sections for these reactions depend on neutrino and antineutrino energies.  Therefore one should measure the difference $\Delta E$ between the average energy of electron antineutrinos and neutrinos,
\begin{equation}
\Delta E=\langle E_{\bar\nu_e}\rangle - \langle E_{\nu_e} \rangle\, .
\end{equation}
If $\Delta E$ is large, the wind can be neutron rich.  However if $\Delta E$ is small, the wind will be proton rich and this is likely a serious problem for r-process nucleosynthesis in the wind \cite{rpocessproblems}.  The wind is the preferred r-process site because, for example, it is a known explosive environment that occurs relatively often so it can easily supply enough r-process material.  However, present simulations of the wind do not have a large enough ratio of free neutrons to seed nuclei.  One can increase this ratio by lowering $Y_p$ (and increasing the number of neutrons) or by increasing the entropy (and destroying some of the seed nuclei).  

Neutrino oscillations can impact $\Delta E$ and or its interpretation.  Oscillations that take place between the neutrino-sphere and the wind can change $\Delta E$ and impact nucleosynthesis \cite{nuoscrprocess}.  However this would not change the interpretation of the measured $\Delta E$ in terms of the composition of the wind.  For example, unexpected neutrino oscillations might raise $\Delta E$ and this could solve many of the problems for r-process nucleosynthesis.  Alternatively, oscillations occurring between the wind and the detectors would not change the nucleosynthesis, but they could complicate the interpretation of $\Delta E$ in terms of the composition of the wind.  These oscillations could occur in the SN envelope, in space between the SN and earth, or in the earth on the way to the detector(s).  We may be able to constrain these oscillations from calculations, or by measuring other neutrino properties such as the energy difference between mu and tau neutrinos and electron anti-neutrinos.  Note that neutrino oscillations in SN can be complicated because of collective effects from neutrino-neutrino interactions \cite{nuosc}.

Finally, calculations of r-process nucleosynthesis yields also require important nuclear structure input including masses of neutron rich nuclei, beta decay half lives, and neutron capture cross sections \cite{rprocess}.  New radioactive beam accelerators, such as FRIB, can directly produce some of the neutron rich nuclei involved in the r-process and provide important nuclear structure data.  However this nuclear data may not, by itself, directly address important astrophysical questions involving the site of the r-process and the source of the neutrons.    

In any case, measuring $\Delta E$, during the next galactic SN, should provide an important diagnostic on conditions in the neutrino driven wind.  At present the site of the r-process is unknown \cite{qian}.  One alternative site for the r-process is tidally ejected material during neutron star mergers, see for example \cite{rprocessmergers}\cite{rprocessmergers2}\cite{rprocessmergers3}.  If this is the site, there may be an observable electromagnetic signal associated with the radioactive decay of r-process nuclei, see for example \cite{Metzger}.  The Spaniards started a quest to find the city of Eldorado.  Perhaps with SN neutrinos or electromagnetic observations of neutron star mergers, we can end this quest and find the source of the chemical element gold.

\subsection{Equation of state for supernova simulations}
Supernova simulations need both neutrino opacities and the equation of state (EOS) of neutron rich matter.  Neutrino opacities describe neutrino interactions in the medium and they should be consistent with the EOS, see for example \cite{reddy}\cite{cjhap}\cite{cjholdions}.  The EOS gives the pressure $P$ as a function of density $n$, temperature $T$, and proton fraction $Y_p$.  The EOS is needed over a very large range of $n$, $T$, and $Y_p$ and must be thermodynamically consistent.  This insures that various partial derivatives of thermodynamic quantities are appropriately related, the first law of thermodynamics is satisfied, and entropy is conserved during adiabatic compression.  Although there are many calculations of the EOS, very few are available over the whole $n$, $T$, and $Y_p$ range in a consistent fashion.  As a result, almost all realistic SN simulations use either the Lattimer Swesty (LS) \cite{LS} or H. Shen et al. (HS) \cite{Hshen} EOS.  A very simple liquid drop model is used for the LS EOS at low density and a Skyrme interaction at high density.  In between, LS assumes a first order liquid-vapor phase transition.  This EOS neglects many nuclear structure features including shell effects, neutron skins, and nuclear pasta.  The HS EOS is based on a relativistic mean field interaction, in a Thomas Fermi and variational approximation.  Although this EOS includes neutron skins, it also neglects shell effects and nuclear pasta.  More recently, Hempel and Schaffner-Bielich have developed an EOS based on a nuclear statistical model \cite{Hempel}.  Although this EOS includes shell effects, it neglects many interactions between nuclei.  

\begin{figure}[htbp]
 \centering
 \includegraphics[width=3.5in,angle=0]{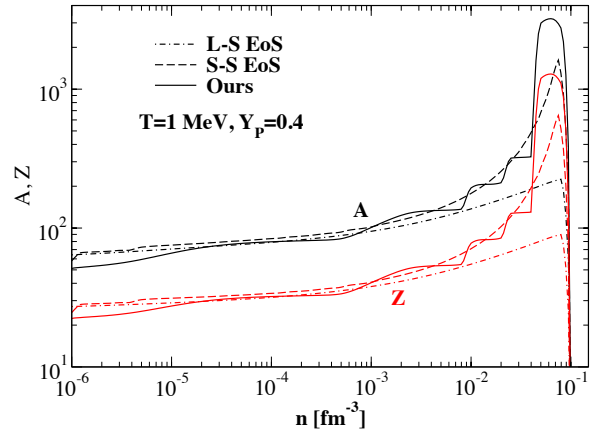}

\caption{Average mass number $A$ (black upper curves) and atomic number $Z$ (red lower curves) of heavy nuclei for our EOS (solid) , Lattimer-Swesty's (dot-dashed) and H. Shen et al's
EOS (dashed) for T = 1 MeV,  $Y_p$ = 0.4 as a function of baryon density $n$ \cite{eos3}.}\label{Fig4}
\end{figure}

We have recently developed new equations of state for use in simulations of SN and neutron star mergers \cite{eos3,eos4}.  At low densities we use a Virial expansion with nucleons, alphas, and thousands of species of heavy nuclei \cite{eos2}.  At high densities we use relativistic mean field calculations in a spherical Wigner-Seitz approximation \cite{eos1}.  Generating the EOS table for a large range of temperatures $T=0$ to 80 MeV, densities $n=10^{-8}$ to 1.6 fm$^{-3}$, and proton fractions $Y_p$ took over 100,000 CPU hours \cite{eos3}.  Our EOS is exact at low densities and contains detailed composition information for calculating neutrino interactions.  In Fig. \ref{Fig4} we show the average mass number $A$ and charge number $Z$ of heavy nuclei as a function of density.  We find plateaus in $A$ or $Z$ for our EOS from shell effects that are absent in either the LS or HS EOS.  Furthermore, there are large differences in composition at higher densities.  We find significantly larger nuclei than in the LS EOS and this may be important for neutrino interactions.

Note that the positions of the plateaus in Fig. \ref{Fig4} correspond to closed shells (``magic numbers'').  These magic numbers are known for the slightly neutron rich conditions $Y_p=0.4$ in Fig. \ref{Fig4}.  However, one also needs the magic numbers for much more neutron rich conditions where they are not presently known.  For example, it was recently shown that $^{24}$O with 8 protons and 16 neutrons may be a doubly closed shell nucleus \cite{O24}.  Experiments with very neutron rich radioactive beams should help determine the position of other shell closures for lower $Y_p$.  In the near future, simulations of core collapse SN, neutron star mergers, and black hole formation using our EOSs, the existing LS and HS EOSs and other new EOSs should allow one to identify features of astrophysical simulations that are sensitive to the EOS and properties of neutron rich matter.  

To conclude this section, the next galactic supernova should be very exciting.  We expect rich electromagnetic signals, unless our view is blocked by large amounts of dust.  Gravitational wave observations of the SN will provide very useful information, see for example \cite{SNGW}.  Combining this with the neutrino signals will likely lead to tremendous advances in our understanding of the dynamics of SN explosions, neutrinos and neutrino oscillations, nucleosynthesis, the formation of neutron stars, and the properties of neutron rich matter.

\section{Conclusions: neutron rich matter}
\label{sec.conclusions}
Neutron rich matter is at the heart of many fundamental questions in Nuclear
Physics and Astrophysics. What are the high density phases of QCD? Where did the chemical elements come from? What is the structure of many compact and energetic objects in the heavens, and what determines their electromagnetic, neutrino, and gravitational-wave radiations? Moreover, neutron rich matter is being studied with an extraordinary variety of new tools such as Facility for Rare Isotope Beams (FRIB) and the Laser Interferometer Gravitational Wave Observatory (LIGO). 

We described the Lead Radius Experiment (PREX) that uses parity violating electron scattering to measure the neutron radius in $^{208}$Pb. This has important implications for neutron stars and their crusts. We discussed X-ray observations of neutron star radii that also have important implications for neutron rich matter.  Gravitational waves (GW) from sources such as neutron star mergers, rotating neutron star mountains, and collective r-mode oscillations open a new window on neutron rich matter.  Using large scale molecular dynamics simulations, we found neutron star crust to be the strongest material known, some 10 billion times stronger than steel.  It can support mountains on rotating neutron stars large enough to generate detectable gravitational waves.  Finally, neutrinos from core collapse supernovae (SN) provide another, qualitatively different probe of neutron rich matter.  Neutrinos come from the neutrino-sphere that is a low density warm gas phase of neutron rich matter.  Neutrino-sphere conditions can be simulated in the laboratory with heavy ion collisions.  Observations of neutrinos can probe nucleosyntheses in SN while SN simulations depend on the equation of state (EOS) of neutron rich matter.  We discussed a new EOS based on virial and relativistic mean field calculations.   

In conclusion, multi-messenger astronomy is based on the widely held belief that combining astronomical observations using photons, gravitational waves, and neutrinos will fundamentally advance our knowledge of compact and energetic objects in the heavens.  Compact objects such as neutron stars are, in fact, giant nuclei, even if they are an extraordinary 18 orders of magnitude larger than a $^{208}$Pb nucleus. Nevertheless, both in the laboratory and in Astrophysics, these objects are made of the same neutrons, that undergo the same strong interactions, and have the same equation of state.  A measurement in one domain, be it Astrophysics or the laboratory,  can have important implications in the other domain.  Therefore, in this colloquium we generalized multi-messenger astronomy  and discussed multi-messenger observations of neutron rich matter.   We believe that combing astronomical observations using photons, GW, and neutrinos, with laboratory experiments on nuclei, heavy ion collisions, and radioactive beams will fundamentally advance our knowledge of the heavens, the dense phases of QCD, the origin of the elements, and of neutron rich matter.

\section*{Acknowledgments}

This work was done in collaboration with many people including D. K. Berry, E. F. Brown, K. Kadau, J. Piekarewicz, and graduate students Liliana Caballero, Helber Dusan, Joe Hughto, Justin Mason, Andre Schneider and Gang Shen.  This work was supported in part by DOE grant DE-FG02-87ER40365 and by the National Science Foundation, TeraGrid grant TG-AST100014.

\end{document}